\documentclass[useAMS,usenatbib,usegraphicx]{mn2e}
\usepackage{mathtools}
\usepackage{amssymb}
\usepackage{xspace}
\usepackage{upgreek}

\newcommand{\wmap}{\emph{WMAP}\xspace}
\newcommand{\planck}{\emph{Planck}\xspace}

\newcommand{\xPI}{$\mathrm{x}PI$}

\def\ltsim{\ifmmode\stackrel{<}{_{\sim}}\else$\stackrel{<}{_{\sim}}$\fi}
\def\gtsim{\ifmmode\stackrel{>}{_{\sim}}\else$\stackrel{>}{_{\sim}}$\fi}

\newcommand{\beq}{\begin{equation}}
\newcommand{\eeq}{\end{equation}}

\title[Polarised Synchrotron and Thermal Dust Emission in the Plane]{Comparing Polarised
  Synchrotron and Thermal Dust Emission in the Galactic Plane}

\author[T. R. Jaffe et al.]{T. R. Jaffe
  $^{1,2}$\thanks{E-mail: 
    tess.jaffe@irap.omp.eu (TRJ); 
    katia.ferriere@irap.omp.eu (KMF);  
    Anthony.Banday@irap.omp.eu (AJB);
    aws@mpe.mpg.de (AWS); 
    eorlando@stanford.edu (EO);
    macias@lpsc.in2p3.fr (JFMP); 
    lfauvet@rssd.esa.int (LF); 
    celine.combet@lpsc.in2p3.fr (CC);
    edith.falgarone@lra.ens.fr (EF); 
} 
  K. M. Ferri\`ere$^{1,2}$\footnotemark[1],
  A. J. Banday$^{1,2}$\footnotemark[1], 
  A. W. Strong$^{3}$\footnotemark[1],\newauthor
  E. Orlando$^{4}$\footnotemark[1],
  J.~F. Mac\'ias-P\'erez$^{5}$\footnotemark[1],
  L. Fauvet$^{6}$\footnotemark[1],
  C. Combet$^{5}$\footnotemark[1],
  E. Falgarone $^{7}$\footnotemark[1]
\\
$^{1}$ Universit\'e de Toulouse; UPS-OMP; IRAP;  Toulouse, France\\
$^{2}$ CNRS; IRAP; 9 Av. colonel Roche, BP 44346, F-31028 Toulouse cedex 4, France\\
$^{3}$Max-Planck-Institut f\"ur Extraterrestrische Physik, Postfach 1312, D-85741 Garching, Germany \\
$^4$W.W. Hansen Experimental Physics Laboratory, Kavli Institute for Particle Astrophysics and Cosmology,  \\ 
\hspace{1cm}Stanford University, Stanford, CA 94305, USA\\
$^5$LPSC, Universit\'e Joseph Fourier Grenoble 1, CNRS/IN2P3, Institut
National Polytechnique de Grenoble, \\ 
\hspace{1cm}53 avenue des Martyrs, 38026 Grenoble cedex, France\\
$^6$European Space Agency, Space Science Department, Keplerlaan 1,
2200AG Noordwijck, The Netherlands \\
$^7$LERMA/LRA, Ecole Normale Sup\'erieure \& Observatoire de Paris, CNRS, 24, rue Lhomond, 75005 Paris, France \\
}
\begin{document}

\date{}

\pagerange{\pageref{firstpage}--\pageref{lastpage}} \pubyear{}

\maketitle

\label{firstpage}

\begin{abstract}
  As the next step toward an improved large scale Galactic magnetic
  field model, we present a simple comparison of polarised synchrotron
  and thermal dust emission on the Galactic plane.  We find that the
  field configuration in our previous model that reproduces the
  polarised synchrotron is not compatible with the \wmap 94~GHz
  polarised emission data.  In particular, the high degree of dust
  polarisation in the outer Galaxy ($90\degr<\ell <270\degr$) implies
  that the fields in the dust-emitting regions are more ordered than
  the average of synchrotron-emitting regions. This new dust
  information allows us to constrain the spatial mixing of the
  coherent and random magnetic field components in the outer Galaxy.
  The inner Galaxy differs in polarisation degree and apparently
  requires a more complicated scenario than our current model.  In the
  scenario that each interstellar component (including fields and now
  dust) follows a spiral arm modulation, as observed in external galaxies,
  the changing degree of ordering of the fields in dust-emitting
  regions may imply that the dust arms and the field component
  arms are shifted as a varying function of Galacto-centric radius.
  We discuss the implications for how the spiral arm compression
  affects the various components of the magnetised interstellar medium
  but conclude that improved data such as that expected from the
  \planck satellite will be required for a thorough analysis.
\end{abstract}


\section{Introduction}

Determining the large-scale structure of the Galactic magnetic fields
is a longstanding and difficult task.  The fields are important to a
variety of astrophysical phenomena from star formation to cosmic ray
propagation, but there are few ways of observing them directly.  Radio
astronomy is the primary way that we study magnetic fields, but any
analysis in only one domain is subject to a variety of degeneracies
that cannot be disentangled without several complementary measurements
and a multi-wavelength approach.

\citet{jaffe:2010,jaffe:2011} model the large-scale Galactic magnetic
fields in the plane using a combination of three observables: the
total and polarised intensity of diffuse synchrotron emission, and
Faraday rotation measures (RMs) of extra-galactic sources.  Other
analyses of the large-scale magnetic fields have been performed using
the full sky (e.g., \citealt{sun:2008},
\citealt{fauvet:2011,fauvet:2012},
\citealt{jansson:2009,jansson:2012}) or the high-latitude sky
\citep{strong:2011}.  We focus on the plane in order to probe through
the full Galactic disk in detail, and we now continue these studies
with the addition of the polarised thermal dust emission.  In addition
to being an independent tracer of the field direction, the dust
emission follows a different spatial distribution and therefore is not
subject to the same uncertainties as previous studies using only
synchrotron emission.

An important aspect of our previous work is to
separate the fields into components defined by their observational
signatures.  As described in \citet{jaffe:2010}, the coherent fields
contribute to both polarised emission as well as to RM, while
isotropic random fields contribute only to the total intensity of the
diffuse emission.  A third component is needed to include the effect
of anisotropy in the turbulence, a component that maintains a common
ordered axis but changes direction stochastically.  This ``ordered
random'' field component contributes to polarised emission but not to
RM.
\footnote{This component is rarely treated by itself, and there is
  some confusion in the literature over how to unambiguously refer to
  it.  \citet{jansson:2012} refer to it as the ``striated'' component.
  For consistency with previous literature (e.g., \citealt{beck:2009})
  and to remove any ambiguity, we prefer the term ``ordered random'',
  noting that simply referring to the ordered field, as we have done in
  the past, might also be mistakenly construed to include the coherent
  component.}  

Since the cosmic ray electron (CRE) distribution is thought to be
fairly smooth (see \S~\ref{sec:constraints} for a discussion), the
contributions to the synchrotron emission from the different magnetic
field components do not depend strongly on their relative spatial
locations.  Our aim is therefore to add the information from the
polarised dust emission from \wmap to study the spatial distribution
of each component.  The geometry of how the polarised emission relates
to the different magnetic field components is identical for diffuse
thermal dust emission.  But the distribution of dust is not as smooth
as the CRE distribution and is known to peak in spiral arms
(see, e.g., \citealt{drimmel:2001}).  This allows us to probe whether
the field components peak in the same ridges or whether, as seen in
external galaxies \citep{beck:2009}, different components may lie
along different arms that may have different pitch angles or be
shifted relative to each other.  Note that \wmap W-band at 94~GHz
primarily probes the relatively cold dust, which has a different
spatial distribution from the hotter dust dominating the infra-red.

We hope also to shed light on the outstanding question of arm versus
inter-arm contrasts in the components of the magnetised interstellar
medium (ISM).  In external galaxies, \citet{fletcher:2011} and
references therein do not find the contrast in synchrotron emission
between arms and inter-arm regions as predicted by the scenario of
compression in the spiral arm shocks.  It remains unclear if this is
due to the limited resolution of the current observations in external
galaxies.  Our method comparing the complementary
observables of synchrotron, dust, and RMs along with a variety of
external constraints allows us to ask similar questions from within
the plane of the Milky Way.


We test a model of the magnetised ISM where each component contains a
spiral arm structure but the arm peaks are not necessarily spatially
coincident.  In particular, we address the question of how to
reproduce the level of dust polarisation in the outer Galaxy (defined
as $90\degr<\ell <270\degr$, i.e. quadrants two and three).  This
region is dominated by the Perseus arm that passes outward of the
Sun's position.  The outer Galaxy is therefore ideal to model
unambiguously the arms of the different components, since we are
looking through one inter-arm region and one arm ridge, beyond which
is expected to be very little ISM.  The inner Galaxy (defined as
$-90\degr<\ell <90\degr$, i.e. quadrants one and four) is more
complicated, and there are already indications that the large-scale
field structure is not a simple spiral (\citealt{vaneck:2011}).

Our simple comparison of polarised synchrotron and dust emission in
the outer Galaxy allows us to study the relative positions of the arms
as well as the arm versus inter-arm contrasts.  This is analogous to
studies such as \citet{patrikeev:2006} or \citet{fletcher:2011}
perform on M51 but with a very different perspective from within the
disk.  This is important to understand the relationship between the
magnetic fields and other components of the ISM and the effect of
spiral arm compression on all components.  We can test the scenario
where gas, dust, and fields are compressed in the spiral arm
shock, which predicts strongly ordered fields in the dust-emitting
regions.  We further discuss the implications of these results for the
inner Galaxy and outline future work.

We describe the data we use in \S~\ref{sec:data} and the modelling and
simulation methods in \S~\ref{sec:methods}.  We then discuss the
inputs and existing constraints on the various components of the
magnetised ISM in \S~\ref{sec:constraints}.  We show the results of our
simulations in \S~\ref{sec:results} and discuss the implications in
\S~\ref{sec:discussion}, along with outlining the further work these
results point toward.

\section{Data}\label{sec:data}

Following our previous work, we use a combination of low-frequency
radio as well as microwave data, in total and polarised intensity, and
examine the profiles in longitude of each frequency along the Galactic
plane.  

\subsection{Radio and low-frequency microwave datasets}

The datasets and references are summarised in Table~\ref{tab:datasets}.
We use the 408~MHz survey by \citet{haslam:1982}, the \wmap 23~GHz
polarisation map from \citet{jarosik:2011}, and three catalogues of RMs,
the Southern and Canadian Galactic Plane Surveys (SGPS and CGPS) of
\citet{brown:2003,brown:2007} and the additional catalogue from
\citet{vaneck:2011}.

\begin{figure*}
%
%
\includegraphics[width=0.7\linewidth]{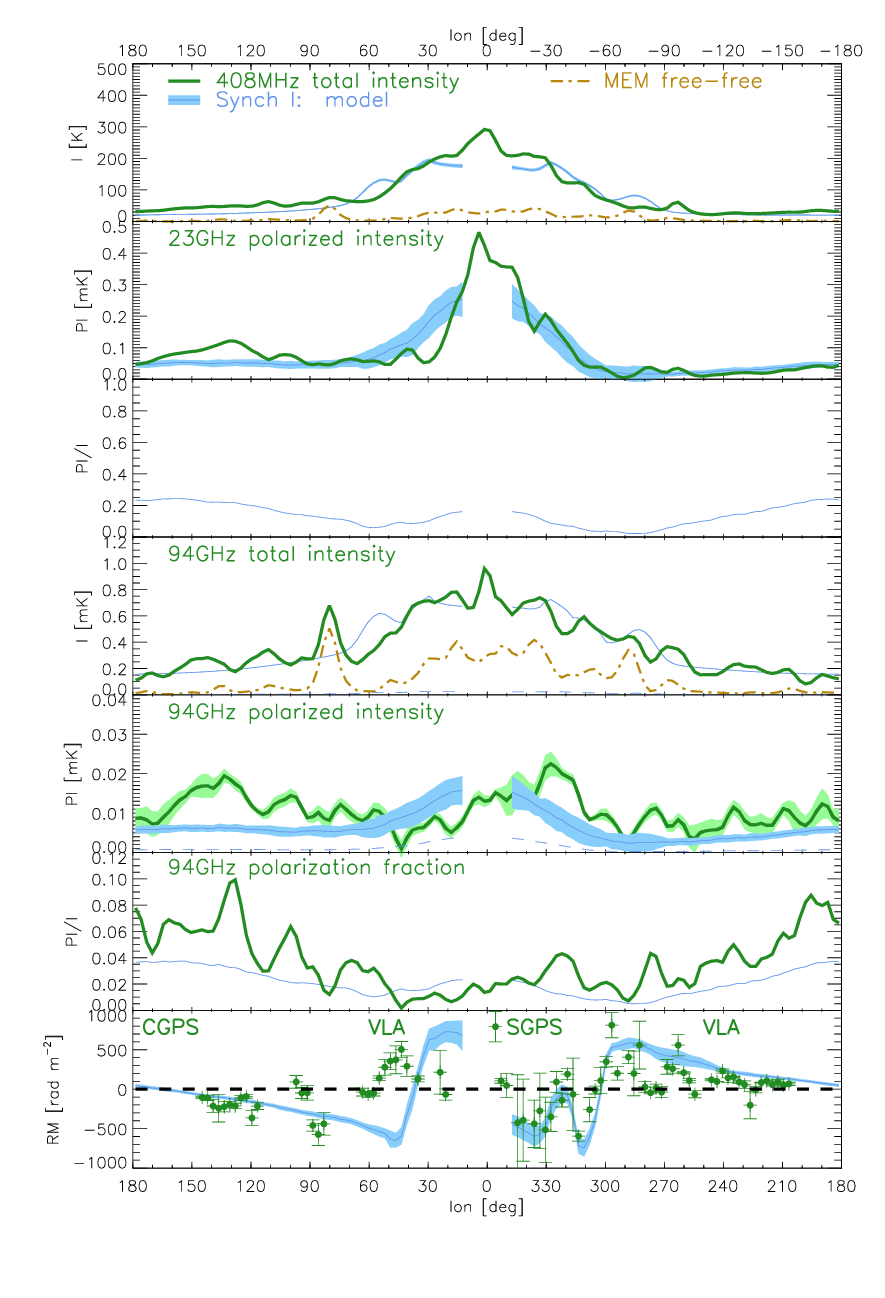}
\caption{Profiles of the observables ({\it thick green lines}) and the slightly
  updated version of our previously fitted model ({\it thin blue lines}) along
  the Galactic plane.  See text in \S~\ref{sec:data}.  In order from the top: total intensity at 408~MHz minus
  free-free estimate; polarised intensity at 23~GHz; model synchrotron
  polarisation fraction; total intensity at 94~GHz minus free-free
  estimate; polarised intensity at 94~GHz and (as the {\it pale green
    filled region}) its uncertainty (see text in
  \S~\ref{sec:data94}); dust polarisation fraction; Faraday RM at
  1.4~GHz.  The shaded regions show the Galactic variance for the
  model, i.e. the expected variation due to the turbulence.  The {\it dashed
  line} in the dust panels shows the modelled synchrotron contribution
  at 94~GHz, which is small but not negligible.  The {\it dot-dashed
    gold line} shows the free-free estimate compared to the
  total intensity curves.}
\label{fig:profile_previous}
\end{figure*}

Each of the emission maps in HEALPix\footnote{{\tt http://healpix.jpl.nasa.gov}}
\citep{healpix} format is smoothed to a common low resolution of FWHM
$6\degr$ and downgraded to $N_\mathrm{side}=32$.  The RM catalogues are
averaged into bins at the same resolution and assigned error bars based on the
RMS variation of the individual measures within the bin.  The profiles in longitude along the
plane are shown in Fig.~\ref{fig:profile_previous} along with a
comparison to an updated form of our previous model \citep{jaffe:2011}
as described in \S~\ref{sec:modeling}.  (Note that we do not attempt
to fit the region within 10$\degr$ of the Galactic centre, and the
model is not plotted there.)

In order to use the low-frequency radio observations in total
intensity on the plane, we apply a correction to remove most of the
free-free emission.  As described in \citet{jaffe:2011}, we find that
the \wmap MEM free-free solution at 33~GHz appears to overestimate the
thermal contribution.  We found, using simple cross-correlations, that
a normalization factor of 0.8 appears to give a good correction to the
radio frequencies, and we use it in the dust band total intensity at
94~GHz as well.  In both cases, the template is extrapolated to each
corrected frequency with the spectrum given in
\citet{dickinson:2003}.  This component has been subtracted from the
total intensity profiles in green in Fig.~\ref{fig:profile_previous}
and is also shown in orange for comparison.

Note that the polarised intensity, $PI=\sqrt{Q^2+U^2}$ plotted in
Fig.~\ref{fig:profile_previous} is subject to a noise bias due to the
squaring of the noise on $Q$ and $U$.  In the case of the synchrotron
polarisation on the plane, the polarisation bias is too small to see.
But for the dust polarisation at 94~GHz, this bias is significant.  To
reduce it, we combine the data from different \wmap years into two
independent maps each of $Q$ and $U$, and then construct
$\mathrm{x}PI=\sqrt{Q_1 Q_2 + U_1 U_2}$.  Since the noise contributions in $Q_1$ and $Q_2$ are
independent, this cross estimator of the polarised intensity has
little noise bias and roughly equivalent uncertainties.  A better estimator of $PI$
would depend on the detailed properties of the detectors, and for the
quantitative analyses in our previous work we used $Q$ and $U$
separately, but for the qualitative analysis presented here, this
$\mathrm{x}PI$ is quite informative.

\begin{table*}
\begin{tabular}{lcccc}
\hline
Survey & Frequency [GHz] & Original resolution & Coverage & Reference \\
\hline
Haslam & 0.408 & $\sim$51 arcmin & Full sky & \citet{haslam:1982}\\
\wmap 7-year & 22.8 & 53 arcmin & Full Sky & \citet{jarosik:2011} \\
\wmap 7-year & 94 & 13 arcmin & Full Sky & \citet{jarosik:2011} \\
CGPS RMs & 1.4 & 1 arcmin & part of Northern plane;  see Fig.~\ref{fig:profile_previous} & \citet{brown:2003} \\
SGPS RMs & 1.4 & 100 arcsec & part of Southern plane;  see Fig.~\ref{fig:profile_previous} & \citet{brown:2007} \\
VLA RMs & 1.4 & 50 arcsec & part of plane;  see Fig.~\ref{fig:profile_previous} & \citet{vaneck:2011}\\
\end{tabular}
\caption{ The datasets used in this analysis.  The RMs are the extra-galactic sources only from the Canadian Galactic Plane survey (CGPS), the Southern Galactic Plane Survey (SGPS) and the van~Eck et al. survey using the Very Large Array (VLA).}
\label{tab:datasets}
\end{table*}

\subsection{\wmap 94~GHz polarisation}\label{sec:data94}

The new dataset we include in this work is the \wmap 94~GHz frequency
that, in polarisation, is dominated by thermal dust emission.  As can be seen in
Fig.~\ref{fig:profile_previous}, the free-free emission correction is
also necessary on the plane, but unlike at 23~GHz, it does not
dominate the total intensity.

The calibration and signal-to-noise ratio of the 94~GHz polarisation
is more of a concern, however, since the dust emission is more weakly
polarised than synchrotron emission.  As noted above, the noise bias
is significant at 94~GHz, which we overcome by plotting
\xPI.  We can also characterise the noise and calibration
uncertainties by comparison of the maps for individual years.  We use
the variance from year to year in $Q$ and $U$ to estimate the uncertainty
on \xPI, which then includes the effects of both the noise and the
systematic uncertainties that are of that time-scale.  These
uncertainties are plotted as the dotted green lines around the green
\xPI profile in Fig.~\ref{fig:profile_previous}.  \citet{jarosik:2011}
single out the W4 difference assembly as particularly prone to
systematic artefacts, and we find that including it does significantly
increase our estimate of the uncertainties.  We therefore follow
Jarosik et al. and omit W4 from the analysis.

The S/N apparent in Fig.~\ref{fig:profile_previous} is somewhat low
for performing quantitative fits and parameter exploration
but is sufficient for the simpler qualitative analysis we present
here.  For a thorough exploration of the parameter space, we will need
better data such as that expected from the \planck \footnote{{\tt
    http://planck.esa.int}} mission \citep{planck:2011}.

\section{Methods}\label{sec:methods}

The methods and tools used in this work are described in
\citet{jaffe:2010}, \citet{jaffe:2011}, and \citet{fauvet:2011}.  In
short, we have adopted models for each of the physical components of
the ISM (thermal electrons, magnetic fields, dust, and CREs) and then
integrated through the Galactic plane to simulate observable polarised
emission and Faraday RM for comparison to the data.  

As discussed in \S~\ref{sec:data94}, the \wmap 94~GHz polarisation
data do not have a high enough S/N for a quantitative analysis such as
the MCMC parameter exploration.  So for this work, we simply compare
the profiles and draw qualitative conclusions from the
morphology.  

\subsection{Models of the Galaxy components}\label{sec:modeling}

We use a magnetic field model based on a spiral arm compression as
described in \citet{jaffe:2010}.  We have improved this model somewhat
to simplify the parametrization and to make the field continuous and
closer to divergence-free.  See Appendix~\ref{sec:appendix} for the
modified parametrization.  The principle differences are: to remove
the extra complexity of the radial profile in the innermost few kpc of
the Galaxy that originally followed \citet{broadbent:1989} but is
excluded in this analysis; to define the transition between arms
continuously (rather than to jump discontinuously in the inter-arm
region as in the previous parametrization); to vary the width of each
arm with Galacto-centric radius in inverse proportion to the coherent
field strength.  Remember that the coherent field is assumed
parallel to the arm, so that the arm corresponds to a magnetic flux
tube. This last modification, which conserves magnetic flux,
maintains a roughly divergence-free field.
%
%
We also allow the spiral arm patterns for the different
components (which are otherwise similar) to vary independently in
orientation about the Galactic north-south pole.  In other words, the
components' arms can be shifted relative to each other by a constant
angle.

As in \citet{jaffe:2011}, we use the ``NE2001'' thermal electron
distribution model of \citet{cordes:2002}.  

We also use the {\sc galprop}\footnote{{\tt http://galprop.stanford.edu/}}  cosmic ray propagation code with a model
based on the ``z04LMPDS'' from \cite{strong:2010}
\footnote{The z04LMPDS was modified in Jaffe et al. (2011) to reduce
  the grid resolution for speed, to disable the irrelevant nucleon
  propagations, and to change the magnetic field used for the
  synchrotron losses.  Most of the parameters for z04LMPDS are defined in
  table~1 of \citet{strong:2010} with the exception of the CR source
  distribution.  The formula is that of \citet{lorimer:2006} (Eq.~10)
  but adjusted in \citet{strong:2011} to agree
better with Fermi-LAT gamma rays, namely $B=0.4751$ and $C=2.1657$ 
and constant for R=10-15~kpc.  
%
%
}
and with the fitted
injection spectrum from \citet{jaffe:2011}.  Based on the given
distribution of CRE sources, the code determines the spatial and
spectral distribution of CREs after propagation in the Galaxy's
magnetic fields, in this case our model rather than the {\sc galprop}
model.  The resulting distribution then is taken to compute the
synchrotron emission as described in \S~\ref{sec:simulation}.

We use a dust distribution model based on the same spiral arm
parametrization used for the magnetic fields, which is in fact similar
to that of \citet{drimmel:2001}.  
Our parameters, given in Table~\ref{tab:params_dust}, are adjusted to
match the integrated total intensity profile of the data at 94~GHz.
Fig.~\ref{fig:profile_previous} shows that the model profile matches
well at large scales.

We use the dust polarisation model described in \citet{fauvet:2011},
i.e. with a geometrical factor of $\sin(\alpha)^3$, where $\alpha$ is
the angle between the local field orientation and the line-of-sight
(LOS).  This factor approximates the effects of grain alignment and
projection onto the LOS, but we note that our analysis is not
particularly sensitive to this factor.

\subsection{Simulation of observables}\label{sec:simulation}

As described in \citet{jaffe:2010}, we simulate the diffuse polarised
emission at large scales using the publicly available code {\sc
  hammurabi}\footnote{{\tt
    http://sourceforge.net/projects/hammurabicode/}}
\citep{waelkens:2009}, which simulates observables such as synchrotron
and dust emission in full Stokes parameters and Faraday RMs.
\citet{jaffe:2011} described how we then linked this code to the {\sc
  Galprop} cosmic ray propagation code in order to include high-energy
constraints from both direct cosmic-ray measurements or from diffuse
$\gamma$-ray emission as measured by the Fermi LAT \citep{fermi:2010,
  abdo:2009}.  This is important to compare the synchrotron total
intensity at low radio frequencies with polarised intensity at
microwave frequencies.  (Since low frequency polarisation is affected
by Faraday depolarisation and higher frequencies by free-free
contamination on the plane, it is complicated to compare total and
polarised intensities on the plane at the same frequency.)  Parameters
have simply been adjusted manually to match the data as plotted in
profiles along the plane as in Fig.~\ref{fig:profile_previous}.  (A
quantitative parameter exploration will be performed in a future
work.)

Fauvet et al. use {\sc hammurabi} to compute the ``geometric
suppression'' map, i.e. the map of the polarisation fraction, and
multiply it by a total intensity template to scale the polarisation
simulation.  This allows a more accurate modelling, since the intensity
template partly accounts for localised features not in the model.  In
Fig.~\ref{fig:profile_simple_comparison}, we follow the same approach,
using the 94~GHz total intensity data corrected for the free-free
emission as described in \citet{jaffe:2011}.  This is in contrast to
Fig.~\ref{fig:profile_previous}, where we do not use a template but
simply plot the result of the integration through our dust density
model.  In this case, both the data and the simulated total intensity
profiles are plotted for comparison, and their very similar
large-scale characteristics give confidence that the dust
distribution model should be accurate enough for this analysis.

As described in \citet{waelkens:2009}, the simulation resolution in
{\sc hammurabi} is refined with radius as the observed solid angle
encompasses a larger volume.  The result is a resolution that varies
but does not increase significantly with radius.  In our large-scale,
low-resolution simulations, it is roughly 20~pc or larger.  With an
outer scale of turbulence of 100~pc, this is clearly not sensitive to
more than the largest scales of the turbulence (represented by a
Gaussian random field simulation).  Any structure on scales smaller
than this will be modeled only as an average over the ``sub-grid''
effects.  We have tested that increasing the resolution does not
impact our results.  The maps themselves are simulated to match
observations smoothed to 6$\degr$ as described in \S~\ref{sec:data}
above.



\section{Inputs and Constraints}
\label{sec:constraints}

We use our previously studied models and additional information from
other studies that is relevant to the comparison of data and model.
In the following, we explain the choices we have made and the relevant
external constraints.

\subsection{Synchrotron}

In \citet{jaffe:2010,jaffe:2011}, we constrained the ratios of the
magnetic field components (defined above) using synchrotron total and
polarised emission as well as Faraday rotation measure.  For the
current analysis, we retain largely the same model as in
\citet{jaffe:2011} but allow for an azimuthal shift between the spiral
patterns of the different components.  This has little effect on the
integrated synchrotron profiles, since the model CRE distribution is
relatively smooth and varies slowly with Galacto-centric radius.

\subsection{Arm versus inter-arm contrast}\label{sec:constraints_arms}

The theory of shock compression at a spiral arm predicts a gas density
contrast of approximately a factor of 4 in the shocked versus the
pre-shock medium, assuming a strong adiabatic shock (e.g.,
\citealt{landau:1987}).  The components of the magnetic field that are
parallel to the shock are therefore compressed as well and thus
amplified by a factor of 4.  This is the physical motivation for the
creation of the ordered random field from the isotropic random field
and for the amplification of the coherent field, which is parallel to
the same spiral shock front in our model.  

We do not model the shock itself but simple spiral arms with Gaussian
profiles.  This can be thought of as a large-scale approximation to
the modulation of a shock that then re-expands to its pre-shock
strength downstream.  We can define a contrast as the ratio of the
peak in the arm to the minimum between arms.  The original model of
\citet{broadbent:1990} used a contrast of 3.5 for both coherent and
isotropic random field components, which they found reproduced well
the synchrotron total intensity profile, particularly the amplitude of
the emission observed in the direction of spiral arm tangents.  

We use the Broadbent et al. value of 3.5 as a starting value, but in
our model, the strength of the coherent field component in each arm is
allowed to vary as in \citet{jaffe:2010} so as to be consistent with
the RM data.  This contrast then varies for each arm by the factor
$a_i$ given in Table~\ref{tab:params_fields}.  But these values cannot
be compared with the theoretical value of 4 for shocked versus
pre-shock medium without some knowledge of the width of the shocked
region and its re-expansion, since our Gaussian profiles are
effectively a large-scale average.  Note also that the coherent field 
strengths in the arms we choose to match the RM data are degenerate
with the thermal electron distribution, which is not very precisely
known.

The ordered random component increases at the shock as a result of
compression of both any pre-shock ordered random field and of the
pre-shock isotropic random field.  It then decreases back to its
pre-shock strength, due to both re-expansion across the spiral arm and
re-isotropisation by supernova-driven turbulence after star formation
triggered by the shock.  There are also other possible sources of the
ordered random component such as superbubbles or shearing motions.  In
our model, this component has a Gaussian profile that goes to zero
between the arms (and therefore has no meaningful contrast), so this
is clearly only a simple approximation to include a variety of
physical effects.

The isotropic random component itself would remain unchanged through
the shock, only providing the source of an ordered random field
created there.  Our modeled spiral arm modulation of the isotropic
random component therefore has a different origin such as the
supernova-driven turbulence downstream of the shock.  The arm contrast
is therefore not constrained a priori and is left at the Broadbent et
al. value of 3.5.  

The diffuse component of the gas and dust would also be compressed by
a factor of 4 in the shock, but the average gas and dust density is
more strongly compressed due to the formation of dense clouds behind
the shock.  Since the \wmap data at 94~GHz trace primarily the cold
dust component, it in particular will be tracing a more strongly
compressed average density.  For our dust model, we use the same
parametrized spiral arm modulation but with a contrast of 7.  This
value, however, is degenerate with other parameters such as the dust
arm width.

Our adopted contrasts are therefore either unconstrained a priori or
roughly compatible with the loose theoretical constraint from the
shock scenario.  We will show that we can reproduce the degree of
polarisation in the data in the outer Galaxy with our current choices,
but we note that this will not provide a unique solution.  Additional
data on the distribution of both the dust and the fields in the Galaxy
will be needed to create an unambiguous model of the effects of
compression on the ISM in the spiral arms.

\subsection{Cosmic ray arm contrast}\label{sec:constraints_high}


The situation for CREs is somewhat different from that of the magnetic
fields and dust.  Similarly to the thermal gas, CREs can be considered
to be tied to field lines\footnote{ 
Although magnetic field lines are a conceptual tool rather than a
physical reality, it is always possible in ideal MHD to define field
lines in such a way that their perpendicular velocity coincides with
the fluid perpendicular velocity, itself equal to the electric drift
velocity \citet{stern:1966}.  Typical CREs, which also have a
perpendicular velocity equal to the electric drift velocity, can then
be regarded as moving with the field lines.
%
%
%
}, so that they should undergo the same perpendicular
compression as the field.  However, in the realistic situation where
the shock front is not infinite and/or the field has a small component
across the shock, CREs would diffuse along field lines at roughly the
Alfv\'en speed, thereby reducing their post-shock density.  An
additional effect, clearly discussed in \citet{fletcher:2011}, is that
the perpendicular momentum squared of each CRE should increase at the
shock in direct proportion to the field strength -- by virtue of the
conservation of the first adiabatic invariant (e.g.,
\citealt{northrop:1963}).  This would lead to an overall shift of the
CRE energy spectrum towards higher energies, and hence to an
additional increase (typically by the same factor of 4 as for the gas
and the magnetic field) in the number density of CREs with a given
energy, e.g. the characteristic energy for a given radio synchrotron
observation.

From these two different effects on the CREs, each of which may lead
to a factor of 4 compression, the number density of CREs at a given
energy is expected to increase at the spiral-arm shock by a factor
between 4 and 16.  This abrupt increase in density would then be 
smoothed out on the downstream side, on roughly a diffusion length, $L
\sim \sqrt{Dt}$, where $D$ is the diffusion coefficient (usually
assumed to be isotropic for simplicity) and $t$ is
the CRE lifetime.  With $D \sim 3 \times 10^{28}~{\rm cm^2~s^{-1}}$
and $t \sim (10^7 - 10^8)$~yrs for electrons of a few GeV in a field
of a few $\upmu$G, the diffusion length works out to be
$L \sim (1-3)$~kpc.

Another aspect that might affect our analysis is the question of
anisotropic diffusion of CRs from their acceleration sites.
\citet{effenberger:2012}, for example, simulate the propagation of CR
protons including anisotropic diffusion in the presence of a spiral
magnetic field.  They find that if the CR sources are also distributed
along spiral arms, the resulting CR density retains a strong azimuthal
modulation, with an amplitude ratio as large as a factor of
$\simeq 6$ for strongly anisotropic diffusion (as opposed to only
$\simeq 2$ for isotropic diffusion) in the case of 1~GeV protons.
Note that this effect is completely separate from, and not necessarily
spatially coincident with, the CRE modulation due to the spiral-arm
shock itself.

The question of whether the CRE distribution varies from arm to
inter-arm regions is therefore a relevant uncertainty in our analysis.

\citet{ackermann:2011} present a study of the third Galactic
quadrant where comparison of the $\gamma$-ray data with the velocity
cubes in HI allows a separation of the spiral arms from the inter-arm
regions.  Among their results is a measure of the $\gamma$-ray
emissivity in each region.  They find
that the $\gamma$-ray emission properties, when normalised by the
varying gas density, vary little as a function of Galacto-centric
radius.  The small ($\approx 5$ per cent) average difference between the arm
and inter-arm region
implies a slightly lower emissivity per unit gas density in the Perseus arm
compared to the inter-arm region that lies just inward of it.  
Though this is an analysis based on CR nucleon interactions rather than CR electrons, 
there is as yet no evidence in favour of any CR enhancement in the Perseus
arm.
  

Since no enhancement as large as predicted has so far been observed
either toward our Galaxy's Perseus arm or in M51
\citep{fletcher:2011}, we consider our model of a smooth, azimuthally
symmetric CRE distribution to be sufficiently realistic for this
first-order analysis.  We discuss the small effect a CRE enhancement
in the arm might have on our results in \S~\ref{sec:results}.

\subsection{Dust intrinsic polarisation fraction}\label{sec:constraints_dustpol}

The details of the thermal dust emission process, and in particular
the alignment of the grains with the magnetic field, are not well
understood \citep{draine:2009}.  There are therefore few theoretical
constraints on the intrinsic polarisation fraction of dust, in
contrast to the case of synchrotron, which introduces an additional
uncertainty into the analysis.  Observations of nearby, high-latitude
clouds, which are likely observed at nearly their intrinsic
polarisation, have dust polarisation fractions of up to $\sim$20 per
cent with most much lower
\citep{benoit:2004,ponthieu:2005,fauvet:2011}.

As we will discuss further below, the interesting differences between
our model and the observations include the overall level of dust
polarisation.  We will show that we can address this problem in the
outer Galaxy by changing the spatial distribution of the magnetic
field components relative to each other and to the dust emitting
regions.  In other words, a higher than expected polarisation fraction
can be reproduced by increasing either the intrinsic polarisation or the
degree of ordering in the fields in the dust emitting regions;  the two
questions are degenerate.  Adopting a higher degree of intrinsic
polarisation is therefore the conservative choice that avoids
over-stating the data's ability to constrain the field ordering.

As described in \S~\ref{sec:simulation}, our simulation has a finite
resolution that in this large-scale study is of order 20~pc.  The
number we use as the ``intrinsic'' polarisation fraction is therefore
the assumed average degree of polarization from a region of this size.
Since the average molecular cloud size is of the same scale, this
fraction then represents an average over a cloud.  We might improve
the modeling of the sub-grid effects, i.e. include the turbulence
driven within the clouds with a model that could be significantly
different in both scale and amplitude from what we have modeled for
synchrotron.  This improvement, however, could only decrease the
predicted degree of polarisation from that obtained assuming an
intrinsic polarisation of 30 per cent.

We therefore adopt 30 per cent for the intrinsic polarization fraction
as a conservative upper limit while noting that this value is
degenerate with the question of how ordered the fields are in the dust
emitting regions.

\subsection{External galaxies}\label{sec:constraints_galaxies}


External galaxies have been mapped in both total and polarised
intensity by low-frequency radio observations.  \citet{beck:2009}
reviews the different morphologies seen and in particular, the
different relationships observed between the spiral structure seen in
polarised synchrotron emission and that seen in gas tracers.  A study
of M51 by \citet{fletcher:2011} provides one of the first estimates of
arm versus inter-arm contrasts seen in polarised synchrotron emission.
They find a much lower level of contrast than expected by the scenario
of shock compression by the spiral wave, but they note that it is
unclear whether their observations have sufficient resolution.  Their
limit on the size of the compressed region based on the observed low
level of contrast and the telescope beam is not so small as to rule
out the scenario.

\citet{patrikeev:2006} use a wavelet analysis to find the arm ridges
in each of a set of maps from low-frequency radio polarisation to CO
and infra-red observations.  They find a systematic shift in the
molecular gas versus the mid-infrared in each spiral arm, consistent
with the delay of star formation triggered in the shock.  The
polarised emission is also the component the furthest upstream, as
expected if the magnetic fields are compressed and aligned with the
shock.  

Though these observations do not provide strong constraints on the
relative positions or widths of the arms or on the arm versus
inter-arm contrasts, they do provide evidence for such shifts and for
variations between components' arm geometries.  As such, observations
of external galaxies motivate our attempt to explore these differences
in our own Galaxy.

\section{Results}\label{sec:results}

\subsection{Profiles along the Galactic plane}\label{sec:results_profiles}

Following \citet{jaffe:2010}, we examine profiles of the observables
along the Galactic plane as shown in Fig.~\ref{fig:profile_previous}.  The
model in blue is simply an updated form of our previous model from
\citet{jaffe:2011}.  The RMs constrain the coherent field
intensity in each arm, while the total and polarised
synchrotron intensities constrain the amounts of isotropic and
ordered random components.  


We have added the dust emission as described in \S~\ref{sec:modeling}
and see that the model (blue curve in Fig.~\ref{fig:profile_previous})
does not fit the dust polarised emission data at all, with two main
problems.  Firstly, the shape of the profile is not correct, with
proportionately too much polarised emission in the inner Galaxy
compared to the outer Galaxy.  Secondly, in the outer Galaxy, the
polarisation fraction is too low.


It is apparent from the RM panel of Fig.~\ref{fig:profile_previous}
that the data from \citet{vaneck:2011} do not match our model.  The
geometry of our model was originally constrained in \citet{jaffe:2010}
before these data were available.  It is clear that though the
amplitude of the closest peak in our RM model roughly matches the
van~Eck et al. data, its longitude does not.  We note that simply
changing the orientations of the spiral arms, their pitch angles
(currently $-11.5\degr$), or the radius of the inner Galaxy ring does
not solve the problem.  Something more complicated than simple
logarithmic spirals and an annular ring will be needed, but the
average RM and its large scale variations predicted by our model are
sufficiently accurate to fix the strength of the coherent field
component.

It must also be noted that there is significant uncertainty regarding the
distance to the Perseus arm in the outer Galaxy.  The spiral arm model
used in the NE2001 model places it roughly 2-3~kpc away toward the
anti-centre, while the analysis of velocity data presented in both
\citet{abdo:2010} and \citet{ackermann:2011} imply nearly twice
the distance.  There is a large degree of uncertainty in both 
estimates, but our analysis is not significantly affected by this.
Our results are dependent only on the {\it relative} positions of the
different components, not their absolute distances, and can therefore be
re-scaled to match a corrected distance determination.

\subsection{Outer Galaxy}\label{sec:results_outer}

For these simulations, we have assumed an intrinsic polarisation
fraction of the dust of 30 per cent.  As discussed in
Section~\ref{sec:constraints_dustpol}, 30 per cent is at the high end of what
is considered plausible and significantly higher than what is usually
assumed.  By using 30 per cent, we are therefore being conservative, since a
more commonly assumed 15 per cent makes the discrepancy between the model and
the data in the outer Galaxy more extreme.  An even higher intrinsic
polarisation fraction might solve the problem in the outer Galaxy but
would be in even more serious conflict with observations.

Nor can we reproduce the observed polarisation fraction by increasing
the average amount of magnetic field ordering, because this would
break the fit to the synchrotron emission.  But since the synchrotron
emission depends on the spatial distribution of CREs, and since these
CREs quickly diffuse away from their acceleration sites (see
discussion in Section~\ref{sec:constraints_high}), the synchrotron
polarisation degree reflects only the average level of coherent and
ordered random fields versus isotropic random fields.  It is not very
sensitive to where those components are spatially located.  Therefore,
we can shift the components relative to one another without much changing
the synchrotron emission profiles.

\begin{figure*}
%
%
%
%
\includegraphics[width=0.7\linewidth]{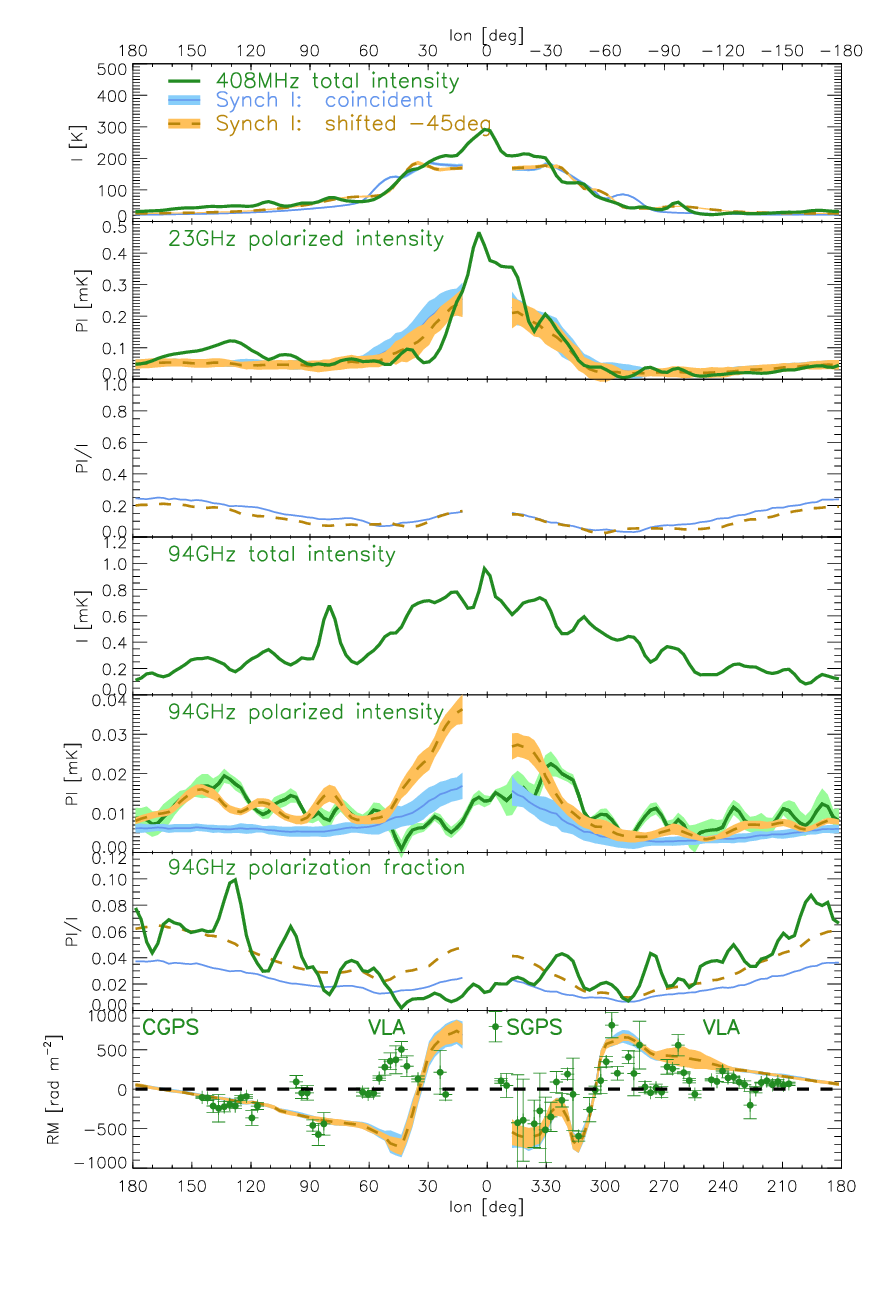}
\caption{Panels as in Fig.~\ref{fig:profile_previous}.  The blue line
  is the model from Fig.~\ref{fig:profile_previous} and has coincident
  arms for all components while the dashed yellow model is that where
  the isotropic random field component has been shifted back 45$\degr$
  in Galactic azimuth, as shown in Fig.~\ref{fig:model}.  See
  \S~\ref{sec:results}.}
\label{fig:profile_simple_comparison}
\end{figure*}

\begin{figure*}
\includegraphics[width=0.7\linewidth]{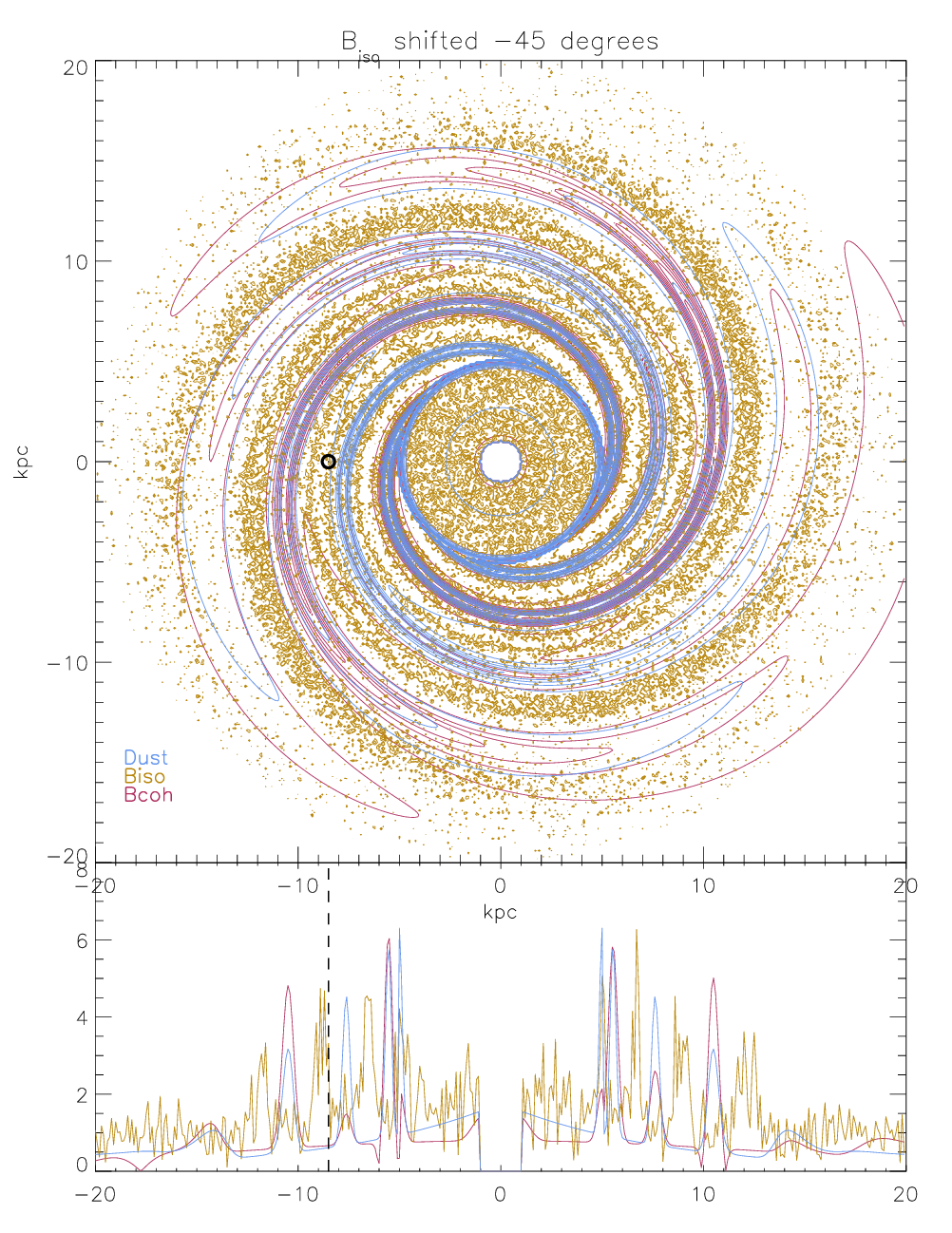}
\caption{Top panel:  contour plot of the Galaxy model seen from above.  The magenta traces the coherent and ordered random magnetic fields.  In blue is the dust distribution.  In gold is the isotropic random field component.  The Sun position is at -8.5~kpc, to the left of the galactic centre.  The lower panel shows the profile on the axis connecting the Sun and the Galactic centre.  The observed profiles along the plane are shown in Fig.~\ref{fig:profile_simple_comparison}.}
\label{fig:model}
\end{figure*}

We show the effect of this with a new model in yellow in
Fig. ~\ref{fig:profile_simple_comparison}, which is identical to the
previous model except for the location of the isotropic random
component.  In the blue curve, as before, all field components as well
as the dust have the peak of the arm in the same place.  But the
dashed yellow curve is the prediction from a model where the coherent
and ordered random fields peak in the same place as the dust but the
isotropic random component has been shifted away by $45\degr$ in
Galacto-centric azimuth.  This places it exactly between two dust
arms, the furthest separation possible, maximising the resulting
polarisation.  The model is shown in Fig.~\ref{fig:model}.  The result
is two very similar profiles for the synchrotron polarisation fraction
but significantly more polarised dust emission, since it is then
emitted in regions of strongly ordered fields.  This then matches far
better the data in the outer Galaxy.

The model parameters have been set to those that best match the outer
Galaxy, particularly the third quadrant (longitudes 180$\degr$ to
270$\degr$), since it is unclear how much the so-called ``Fan'' region
(dominating most of the second quadrant, 90$\degr$ to 180$\degr$),
which remains anomalous, may be due to a local feature.   

\subsection{Inner Galaxy}\label{sec:results_inner}

The shift of the isotropic random field component away from the dust
emitting regions makes the mismatch between model and observations in
the inner Galaxy worse.  The inner Galaxy is also complicated by the
presence of what is often referred to as the ``molecular ring'' and
whose effect on the magnetic field components is not known.
Furthermore, our Galaxy is thought to have a bar (see, e.g.,
\citealt{gerhard:2002}) with again unknown effect on the field
components and on the dust.

The parameters that match the dust emission in the outer Galaxy fail
in the inner Galaxy, and this tension cannot be resolved with the
current parametrization of the fields. Clearly, a more complicated
model that smoothly varies quantities such as the arm positions,
widths, contrasts, and pitch angles from the inner to outer Galaxy
will be necessary to fit the data.

\subsection{Comparison to low-frequency radio polarisation}

Note that the model in Fig.~\ref{fig:model} has the Sun position
(8.5~kpc to the left of centre) just inside the isotropic random field
arm.  This means that looking toward the outer Galaxy, we ought
to see Faraday depolarisation effects in the low-frequency radio data
in that direction.  Though the strength of the depolarisation effect
is unknown and would require some additional modeling to estimate, it
is clear that the effect is the opposite of what is seen, namely that
it is the inner Galaxy where there is significant Faraday
depolarisation and the outer Galaxy only where strong polarisation is
apparent.  See, e.g., fig.~1 of \citet{burigana:2006} showing the
polarised emission at 1.4~GHz.  The implication is that the positions
of the arms in our model may need to be adjusted relative to the
position of the Sun.

We will in future add to our studies the Faraday depolarisation effects seen in the
1.4~GHz polarisation map, since this is important additional data on
the spatial distribution of the turbulent medium.


\section{Discussion}\label{sec:discussion}

We show in Section~\ref{sec:results} how the profiles of polarised
synchrotron and dust emission in the Galactic plane, combined with
previous work to constrain average strengths of the field components
in the Galaxy, require a spatial separation of the different field
components.  It is clear from Fig.~\ref{fig:profile_simple_comparison}
that a constant mixing of the coherent plus ordered random fields with
the isotropic random component is inconsistent with the polarised dust
data in the Galactic plane.  We can also see that the problem is
opposite in the inner and outer Galaxy.
Our current model cannot simultaneously fit both regions, but we can
already draw some conclusions from
Fig.~\ref{fig:profile_simple_comparison}.



\subsection{The shock scenario}

Firstly, we can reproduce the emission in the outer Galaxy with a
model where the isotropic random component peaks in arms that
are clearly separated from both the coherent and ordered random
components and the dust emitting regions.  The synchrotron emission
constrains only the average amount of coherent plus ordered random and
isotropic random magnetic fields in the Galaxy, but does not strongly
depend on where they are located.  External galaxies show indications
of magnetic arms distinct from the gas arms and with morphologies that
are not necessarily the same (see, e.g., \citealt{frick:2000} or \citealt{beck:2009}).  They
can be shifted relative to each other, have different pitch angles, or
cross over the gas arm ridges.  There is clear evidence for such
effects in the case of M51 \citep{patrikeev:2006}.  Our results in the
outer Galaxy support the idea of a shift between components.

The second conclusion we can draw from
Fig.~\ref{fig:profile_simple_comparison} is that the situation is
reversed in the inner Galaxy, where the data show no peak in the
polarised emission as predicted, implying that the fields are less
ordered in the dust emitting regions than in our models.  Our current
model cannot be tuned to simultaneously fit both regions.

We can reproduce the polarisation in the outer Galaxy with a model
that has the peak of dust emission spatially coincident with the
coherent and ordered random fields but offset from the isotropic
random field.  Such a lag can be naturally explained by the idea of a
large-scale shock wave associated with the spiral arm: the shock
compresses the fields so they lie predominantly parallel to the shock,
compresses the gas and dust likewise, and triggers collapse of dust clouds and
subsequently star formation.  The turbulence is then generated in
supernovae downstream of the compression after the stars have had time
to evolve, thus explaining the shift in the isotropic random field
component relative to the other arms.

This scenario is also consistent with the studies by 
\citet{haverkorn:2006b,haverkorn:2008}, which examine the structure
function of Faraday RM fluctuations in directions either tangent to
spiral arms or tangent to inter-arm regions.  They find that in the
{\it inter-arm} regions, the structure function shows a rising
spectrum up to a cut-off scale, consistent with turbulence from a
Kolmogorov cascade of energies from large scale turbulence injected by
supernova remnants.  The arm tangents by contrast do not show such a
spectrum but rather a flat structure function, perhaps implying a much
smaller outer scale of turbulence, small enough that the turn-over on
the structure function is not visible.  This is interpreted by
Haverkorn et al. as an alternative source of turbulence, e.g. HII
regions and stellar winds instead of supernova remnants.  But it is
also what one might expect to see if the larger scale turbulence has
been compressed perpendicular to the shock, i.e. perpendicular to the LOS
when looking tangent to the arms.  Alternatively, it could simply be
that the the star formation is triggered in the shock and only later
evolves into the supernova remnants that inject the turbulence, which
would mean the turbulence is always downstream of the compressed
region.

Since the downstream region is leading the shock inside co-rotation
(where the interstellar gas rotates faster than the spiral pattern)
and lagging behind the shock outside co-rotation, our simple model of
spiral arms being azimuthally shifted as wholes is not accurate.
The value of the co-rotation radius is very uncertain (e.g.,
\citealt{gerhard:2011}), but it could be close to the Galacto-centric
radius of the Sun, in which case inner/outer Galaxy would roughly
coincide with inside/outside co-rotation.  In other words, the
depolarising turbulence would be shifted ahead of the dust arms in
the inner Galaxy, and behind the dust arms in the outer Galaxy.  But
this difference in the shift direction would not change the
polarisation degree.

It should be noted that the dust is emitted mostly in relatively
compact clouds that are not resolved by our simulations.  One can
easily imagine different degrees of ordering in the magnetic fields in
the clouds compared to the diffuse ISM traced by synchrotron emission,
e.g. by turbulence injected at smaller scales.  As discussed in
\S~\ref{sec:constraints_dustpol}, however, our modeling effectively
assumes 30 per cent polarization at the scale of a typical cloud,
which is significantly higher than anything observed.  Furthermore,
there is also a truly diffuse dust component that would trace the
large-scale turbulence seen in synchrotron emission, though the ratio
of such emission to that from relatively compact clouds is unknown.
We therefore find it difficult to envision a plausible alternative
scenario to increase the dust polarization in the outer Galaxy as long
as the dust clouds remain embedded in the same turbulent ISM that is
traced by the synchrotron.

\subsection{Refinements for the inner Galaxy}

There are several possibilities that might explain the difference in
the inner Galaxy.  Broadly speaking, they are either that the
dust-emitting regions have more disordered fields in the inner Galaxy,
or that the intrinsic polarisation fraction of the dust emission
drops.  

We could perhaps model the former by allowing the field components to
have different pitch angles so that the isotropic random component is
coincident with the dust in the inner Galaxy but gradually shifts away
in the outer Galaxy.  This would result in a degree of ordering that
would vary with Galacto-centric radius and might be made to reproduce
the data.  But this would not be consistent with the shock compression
scenario that naturally accounts for the polarisation in the outer
Galaxy, and we currently lack a plausible alternative scenario for
such a morphology.  

There is similarly a question of when the collapsing dust clouds
decouple from the magnetic fields.  There is more star formation in
the inner Galaxy, and the colder dusty regions could be less strongly
coupled to the large-scale ordered fields in the inner Galaxy.  Even
if the clouds remain coupled to the fields while collapsing, the
collapse itself could disrupt the ordering in the fields.  Since the
synchrotron emission comes from a large volume while the dust emission
may be dominated by relatively isolated regions of collapsing clouds,
the dust emitting regions may have more random fields at small scales
embedded in the larger-scale fields traced by synchrotron.  

Alternatively, the truly intrinsic polarisation fraction could drop in
the inner Galaxy due to the different environment (e.g. temperature)
of the dusty regions.  As discussed in
\S~\ref{sec:constraints_dustpol}, the intrinsic polarisation fraction
is not well understood.

We could fold these possibilities into an effective sub-grid
polarisation fraction that varies with Galacto-centric radius.  Our
large-scale analysis cannot distinguish these possibilities without
additional external constraints on small-scale dust polarization
properties, but higher-resolution studies of the fields in individual
clouds may be very informative.


\subsection{Caveats}

Note that the uncertainty in the true intrinsic polarisation fraction
of dust does not affect our analysis in the outer Galaxy.  As
discussed in \S~\ref{sec:constraints_dustpol}, it is very unlikely to
be high enough to account for the observed polarisation without
spatially separating the dust from the isotropic random fields.

Another uncertainty in this analysis is the impact of the possible
modulation of the CRE density due to the two effects discussed in
\S~\ref{sec:constraints_high}, namely the compression and acceleration
of the CREs at the spiral-arm shocks and the distribution of CRE
injection sites.  Both sources of CRE density modulation arise in
spiral arms, but they are not necessarily spatially coincident, as
CREs are presumably injected by supernova remnants in the turbulent
regions that develop downstream from the shocks, away from the
amplified fields. One can speculate that the two effects may result in
a fairly uniform average CRE density across spiral arms, thus
explaining the observed lack of any such modulation in the CRE
density.

In order to reproduce the observed level of synchrotron emission, and
not to result in arm tangents that are more peaked in emission than
observed, any CRE enhancement would have to be offset by a
corresponding lowering of the coherent and ordered random magnetic
fields. In any case, the degree of polarisation of the dust emission
does not depend on the strength of the fields, only on the degree of
ordering in the dusty regions.  So the question of the CRE density is
an interesting one but does not have a large effect on our main
conclusions.

Neither the uncertainty in the intrinsic polarisation fraction of dust
emission nor in the CRE density modulation, therefore, will change our
main conclusion:  that in the outer Galaxy, the dust emitting regions
have more ordered fields than the average as traced by synchrotron
emission.

\section{Conclusions}

We have continued our project to model the large scale Galactic
magnetic fields by adding the polarised thermal dust emission from the
\wmap 94~GHz band.  We find that our model previously fit to
synchrotron and Faraday RMs does not reproduce the observed level of
dust polarisation.  Even with an assumed intrinsic polarisation
fraction of 30 per cent, which is higher than observed, the level of
predicted polarisation is too low in all but the innermost regions of
the Galactic plane.  This implies that the dust emitting regions have
more ordered magnetic fields than average.  We are able to reproduce
the data in the outer Galaxy by modifying the model to spatially
separate the isotropic random field component from the dust emitting
regions and the coherent and ordered random field components.  This
separation is consistent with the scenario of fields and dust
compressed in the spiral arm shocks, with turbulence generated by
supernova remnants downstream.  The same model results in an
over-prediction of the dust polarisation in the inner regions of the
Galaxy, which indicates that a more sophisticated model is necessary.
We will in further work explore models that allow either the spiral
arm characteristics of each component or the dust intrinsic
polarisation degree to vary with Galacto-centric radius.  Our models
would be improved by numerical simulations of galaxies' magnetic
fields that included arm versus inter-arm variations and provided
constraints to break some of our parameter degeneracies.  This project
will be significantly aided by the \planck satellite's more sensitive
observations of polarised dust emission at both large and small
scales.


\section*{Acknowledgements}
We thank D. Alina, J.-P. Bernard for useful discussions.  We acknowledge use of
the HEALPix software \citep{healpix} for some of the results in this
work.  We acknowledge the use of the Legacy Archive for Microwave
Background Data Analysis (LAMBDA). Support for LAMBDA is provided by
the NASA Office of Space Science. 
This research was supported by the Agence Nationale de la Recherche (ANR-08-CEXC-0002-01).

\bibliographystyle{mn2e}
\bibliography{references,refs_custom}

\begin{thebibliography}{}

\bibitem[\protect\citeauthoryear{Abdo et~al.,}{Abdo et~al.}{2009}]{abdo:2009}
Abdo A.~A.  et~al., 2009, Physical Review Letters, 103, 251101

\bibitem[\protect\citeauthoryear{Abdo et~al.,}{Abdo et~al.}{2010}]{abdo:2010}
Abdo A.~A.  et~al., 2010, The Astrophysical Journal, 710, 133

\bibitem[\protect\citeauthoryear{Ackermann et~al.,}{Ackermann
  et~al.}{2010}]{fermi:2010}
Ackermann M.  et~al., 2010, Physical Review D, 82, 92004

\bibitem[\protect\citeauthoryear{Ackermann et~al.,}{Ackermann
  et~al.}{2011}]{ackermann:2011}
Ackermann M.  et~al., 2011, The Astrophysical Journal, 726, 81

\bibitem[\protect\citeauthoryear{Beck}{Beck}{2009}]{beck:2009}
Beck R.,  2009, Proc. IAU Symposium, 259, pp 3--14

\bibitem[\protect\citeauthoryear{Beno{\^\i}t et~al.,}{Beno{\^\i}t
  et~al.}{2004}]{benoit:2004}
Beno{\^\i}t A.  et~al., 2004, Astronomy and Astrophysics, 424, 571

\bibitem[\protect\citeauthoryear{Broadbent}{Broadbent}{1989}]{broadbent:1989}
Broadbent A.,  1989, (PhD thesis, Durham Univ.)

\bibitem[\protect\citeauthoryear{Broadbent, Haslam \& Osborne}{Broadbent
  et~al.}{1990}]{broadbent:1990}
Broadbent A.,  Haslam C. G.~T.,    Osborne J.~L.,  1990, in Proceedings of the
  21st International Cosmic Ray Conference. Volume 3 (OG Sessions). p.~229

\bibitem[\protect\citeauthoryear{Brown, Haverkorn, Gaensler, Taylor, Bizunok,
  McClure-Griffiths, Dickey \& Green}{Brown et~al.}{2007}]{brown:2007}
Brown J.~C.,  Haverkorn M.,  Gaensler B.~M.,  Taylor A.~R.,  Bizunok N.~S.,
  McClure-Griffiths N.~M.,  Dickey J.~M.,    Green A.~J.,  2007, \apj, 663, 258

\bibitem[\protect\citeauthoryear{Brown, Taylor \& Jackel}{Brown
  et~al.}{2003}]{brown:2003}
Brown J.~C.,  Taylor A.~R.,    Jackel B.~J.,  2003, \apjs, 145, 213

\bibitem[\protect\citeauthoryear{Burigana, La~Porta, Reich \& Reich}{Burigana
  et~al.}{2006}]{burigana:2006}
Burigana C.,  La~Porta L.,  Reich P.,    Reich W.,  2006, Astronomische
  Nachrichten, 327, 491

\bibitem[\protect\citeauthoryear{Cordes \& Lazio}{Cordes \&
  Lazio}{2002}]{cordes:2002}
Cordes J.~M.,  Lazio T. J.~W.,  2002, preprint (astro-ph/0207156)

\bibitem[\protect\citeauthoryear{Dickinson, Davies \& Davis}{Dickinson
  et~al.}{2003}]{dickinson:2003}
Dickinson C.,  Davies R.~D.,    Davis R.~J.,  2003, \mnras, 341, 369

\bibitem[\protect\citeauthoryear{Draine \& Fraisse}{Draine \&
  Fraisse}{2009}]{draine:2009}
Draine B.~T.,  Fraisse A.~A.,  2009, The Astrophysical Journal, 696, 1

\bibitem[\protect\citeauthoryear{Drimmel \& Spergel}{Drimmel \&
  Spergel}{2001}]{drimmel:2001}
Drimmel R.,  Spergel D.~N.,  2001, \apj, 556, 181

\bibitem[\protect\citeauthoryear{Effenberger, Fichtner, Scherer \&
  Buesching}{Effenberger et~al.}{2012}]{effenberger:2012}
Effenberger F.,  Fichtner H.,  Scherer K.,    Buesching I.,  2012,
  astro-ph/1210.1423

\bibitem[\protect\citeauthoryear{Fauvet et~al.,}{Fauvet
  et~al.}{2011}]{fauvet:2011}
Fauvet L.  et~al., 2011, Astronomy and Astrophysics, 526, 145

\bibitem[\protect\citeauthoryear{Fauvet, Mac{\'\i}as-P{\'e}rez, Jaffe, Banday,
  D{\'e}sert \& Santos}{Fauvet et~al.}{2012}]{fauvet:2012}
Fauvet L.,  Mac{\'\i}as-P{\'e}rez J.~F.,  Jaffe T.~R.,  Banday A.~J.,
  D{\'e}sert F.~X.,    Santos D.,  2012, Astronomy and Astrophysics, 540, 122

\bibitem[\protect\citeauthoryear{Fletcher, Beck, Shukurov, Berkhuijsen \&
  Horellou}{Fletcher et~al.}{2011}]{fletcher:2011}
Fletcher A.,  Beck R.,  Shukurov A.,  Berkhuijsen E.,    Horellou C.,  2011,
  Monthly Notices of the Royal Astronomical Society, 412, 2396

\bibitem[\protect\citeauthoryear{Frick, Beck, Shukurov, Sokoloff, Ehle \&
  Kamphuis}{Frick et~al.}{2000}]{frick:2000}
Frick P.,  Beck R.,  Shukurov A.,  Sokoloff D.,  Ehle M.,    Kamphuis J.,
  2000, Monthly Notices of the Royal Astronomical Society, 318, 925

\bibitem[\protect\citeauthoryear{Gerhard}{Gerhard}{2002}]{gerhard:2002}
Gerhard O.,  2002, in Proceedings. p.~73

\bibitem[\protect\citeauthoryear{Gerhard}{Gerhard}{2011}]{gerhard:2011}
Gerhard O.,  2011, Memorie della Societa Astronomica Italiana Supplementi, 18,
  185

\bibitem[\protect\citeauthoryear{G{\'o}rski, Hivon, Banday, Wandelt, Hansen,
  Reinecke \& Bartelmann}{G{\'o}rski et~al.}{2005}]{healpix}
G{\'o}rski K.~M.,  Hivon E.,  Banday A.~J.,  Wandelt B.~D.,  Hansen F.~K.,
  Reinecke M.,    Bartelmann M.,  2005, \apj, 622, 759

\bibitem[\protect\citeauthoryear{Han, Ferriere \& Manchester}{Han
  et~al.}{2004}]{han:2004}
Han J.~L.,  Ferriere K.,    Manchester R.~N.,  2004, \apj, 610, 820

\bibitem[\protect\citeauthoryear{Haslam, Stoffel, Salter \& Wilson}{Haslam
  et~al.}{1982}]{haslam:1982}
Haslam C. G.~T.,  Stoffel H.,  Salter C.~J.,    Wilson W.~E.,  1982, \aaps, 47,
  1

\bibitem[\protect\citeauthoryear{Haverkorn, Brown, Gaensler \&
  McClure-Griffiths}{Haverkorn et~al.}{2008}]{haverkorn:2008}
Haverkorn M.,  Brown J.~C.,  Gaensler B.~M.,    McClure-Griffiths N.~M.,  2008,
  The Astrophysical Journal, 680, 362

\bibitem[\protect\citeauthoryear{Haverkorn, Gaensler, Brown, Bizunok,
  McClure-Griffiths, Dickey \& Green}{Haverkorn et~al.}{2006}]{haverkorn:2006b}
Haverkorn M.,  Gaensler B.~M.,  Brown J.~C.,  Bizunok N.~S.,  McClure-Griffiths
  N.~M.,  Dickey J.~M.,    Green A.~J.,  2006, The Astrophysical Journal, 637,
  L33

\bibitem[\protect\citeauthoryear{Jaffe, Banday, Leahy, Leach \& Strong}{Jaffe
  et~al.}{2011}]{jaffe:2011}
Jaffe T.~R.,  Banday A.~J.,  Leahy J.~P.,  Leach S.,    Strong A.~W.,  2011,
  Monthly Notices of the Royal Astronomical Society, 416, 1152

\bibitem[\protect\citeauthoryear{Jaffe, Leahy, Banday, Leach, Lowe \&
  Wilkinson}{Jaffe et~al.}{2010}]{jaffe:2010}
Jaffe T.~R.,  Leahy J.~P.,  Banday A.~J.,  Leach S.~M.,  Lowe S.~R.,
  Wilkinson A.,  2010, Monthly Notices of the Royal Astronomical Society, 401,
  1013

\bibitem[\protect\citeauthoryear{Jansson \& Farrar}{Jansson \&
  Farrar}{2012}]{jansson:2012}
Jansson R.,  Farrar G.~R.,  2012, \apj, 757, 14

\bibitem[\protect\citeauthoryear{Jansson, Farrar, Waelkens \&
  En{\ss}lin}{Jansson et~al.}{2009}]{jansson:2009}
Jansson R.,  Farrar G.~R.,  Waelkens A.~H.,    En{\ss}lin T.~A.,  2009, Journal
  of Cosmology and Astro-Particle Physics, 7, 21

\bibitem[\protect\citeauthoryear{Jarosik et~al.,}{Jarosik
  et~al.}{2011}]{jarosik:2011}
Jarosik N.  et~al., 2011, The Astrophysical Journal Supplement, 192, 14

\bibitem[\protect\citeauthoryear{Landau \& Lifshitz}{Landau \&
  Lifshitz}{1987}]{landau:1987}
Landau Lifshitz 1987, {Fluid Mechanics}.
 Vol. 6

\bibitem[\protect\citeauthoryear{Lorimer et~al.,}{Lorimer
  et~al.}{2006}]{lorimer:2006}
Lorimer D.~R.  et~al., 2006, Monthly Notices of the Royal Astronomical Society,
  372, 777

\bibitem[\protect\citeauthoryear{Northrop}{Northrop}{1963}]{northrop:1963}
Northrop T.,  1963, {The adiabatic motion of charged particles}

\bibitem[\protect\citeauthoryear{Patrikeev, Fletcher, Stepanov, Beck,
  Berkhuijsen, Frick \& Horellou}{Patrikeev et~al.}{2006}]{patrikeev:2006}
Patrikeev I.,  Fletcher A.,  Stepanov R.,  Beck R.,  Berkhuijsen E.~M.,  Frick
  P.,    Horellou C.,  2006, Astronomy and Astrophysics, 458, 441

\bibitem[\protect\citeauthoryear{{Planck Collaboration}}{{Planck
  Collaboration}}{2011}]{planck:2011}
{Planck Collaboration} 2011, \aap, 536, A1

\bibitem[\protect\citeauthoryear{Ponthieu et~al.,}{Ponthieu
  et~al.}{2005}]{ponthieu:2005}
Ponthieu N.  et~al., 2005, Astronomy and Astrophysics, 444, 327

\bibitem[\protect\citeauthoryear{Stern}{Stern}{1966}]{stern:1966}
Stern D.~P.,  1966, Space Science Reviews, 6, 147

\bibitem[\protect\citeauthoryear{Strong, Porter, Digel, J{\'o}hannesson,
  Martin, Moskalenko, Murphy \& Orlando}{Strong et~al.}{2010}]{strong:2010}
Strong A.,  Porter T.,  Digel S.,  J{\'o}hannesson G.,  Martin P.,  Moskalenko
  I.,  Murphy E.,    Orlando E.,  2010, The Astrophysical Journal Letters, 722,
  L58

\bibitem[\protect\citeauthoryear{Strong, Orlando \& Jaffe}{Strong
  et~al.}{2011}]{strong:2011}
Strong A.~W.,  Orlando E.,    Jaffe T.~R.,  2011, Astronomy and Astrophysics,
  534, 54

\bibitem[\protect\citeauthoryear{Sun, Reich, Waelkens \& En{\ss}lin}{Sun
  et~al.}{2008}]{sun:2008}
Sun X.~H.,  Reich W.,  Waelkens A.,    En{\ss}lin T.~A.,  2008, \aap, 477, 573

\bibitem[\protect\citeauthoryear{{van~Eck} et~al.,}{{van~Eck}
  et~al.}{2011}]{vaneck:2011}
{van~Eck} C.  et~al., 2011, \apj, 728, 97

\bibitem[\protect\citeauthoryear{Waelkens, Jaffe, Reinecke, Kitaura \&
  En{\ss}lin}{Waelkens et~al.}{2009}]{waelkens:2009}
Waelkens A.,  Jaffe T.,  Reinecke M.,  Kitaura F.~S.,    En{\ss}lin T.~A.,
  2009, \aap, 495, 697

\end{thebibliography}

\appendix
\section{Model parametrization}
\label{sec:appendix}

The following tables describe the parametrization of the magnetic
field and dust distribution models.  The
difference to previous works is in simplifying the radial profiles
(since we do not fit the innermost region of the Galaxy where the
\cite{broadbent:1989} features have effect) and in making the field
continuous in the transitions between arms.  Each component has an
associated scale height, which is unconstrained by our analysis
confined to the Galactic plane.  But they are given for completeness
and for the small effect they may have within the resolution of
6$\degr$.

\begin{table*}
\begin{center}
\begin{minipage}{\linewidth}
\begin{tabular}{ | l | p{1.5cm} | p{7.5cm} | p{6cm} |}	

\hline
\multicolumn{4}{c}{{\bf Total field} $\mathbf{B}_\mathrm{tot} =
  \mathbf{B}_\mathrm{coh} + \mathbf{B}_\mathrm{iso} +
  \mathbf{B}_\mathrm{ord} $} \\
\hline
 Param. & Default & Equation & Description \\
\hline
\multicolumn{4}{c}{{\bf Axisymmetric spiral coherent magnetic field:}  $\mathbf{B}_\mathrm{ASS}(r,\phi,z)=B(r) B_\mathrm{coh}(z) \mathbf{\hat B}$} \\
\hline
$B_0$ & 1 $\upmu$G & &  Global amplitude normalisation\\

$R_\mathrm{scale}$ & 20 kpc & $B(r)=B_0\exp(-r^2/R_\mathrm{scale}^2)$ & Outer radial profile parameter.  See fig.~3 of \citet{jaffe:2010}. \\

$h_{\mathrm g}$  &  6~kpc &  $B_\mathrm{coh}(z) = \mathrm{sech}^2(z/h_{\mathrm g})$ & Global scale height of the coherent field. \\

$R_\rmn{max}$ & 20 kpc & & Maximum radius, beyond which $|B|=0$\\

$R_\rmn{mol}$  & 5 kpc & $\mathbf{\hat B} = \left\{  \begin{array}{ll}
    (\cos(\phi+\pi/2)\mathbf{\hat e}_x, \sin(\phi+\pi/2)\mathbf{\hat e}_y,0)  &  \mbox{ if $r\le R_\mathrm{mol}$ } \\
    (\sin(\theta_p)\mathbf{\hat e}_r, \cos(\theta_p)\mathbf{\hat e}_\phi, 0)      &  \mbox{ if $r>R_\mathrm{mol}$ } \\ 
\end{array} \right.  $ 
& Field direction circular within ``molecular ring'' and otherwise spiral (expressed in cylindrical coordinates, with the x-axis between the Sun and Galactic centre). \\

$\theta_p$ & -11.5$\degr$ &  &  $\theta_p$ is the pitch angle of field direction and spiral arms \\

\hline
\multicolumn{4}{c}{ {\bf Spiral compression arm coherent magnetic field:} $\mathbf{B}_\mathrm{coh} = \mathbf{B}_\mathrm{ASS} + \sum_i^{N_\mathrm{arms}} \mathbf{B}_{\mathrm{arm},i}$} \\
\hline

$a_i$ & $(3, 0.5, -4, $\linebreak$1.2, -0.8)$ & $\mathbf{B}_{\mathrm{arm},i}=  B(r) a_i \rho_c(d_i)\mathbf{\hat
  B} $ & Amplitude of each of four spiral arms and molecular ring.. \\


&  &  $\rho_c(d)= c(r) B_\mathrm{comp}(z) \exp(-(d/d_0(r))^2)$ &  Additional compression factor relative to background.  $d$ is the distance to the nearest arm in kpc, computed using $r_i(\phi)$ \\

$\phi_{0,i}$ & $10\degr+90\degr i$ & $r_i(\phi)=R_s
\exp\left[(\phi-\phi_{0,i})/\beta\right]$ and $\beta\equiv 1/\tan(\theta_p)$ & $r(\phi)$ gives arm radius at given azimuth, and $\phi_{0,i}$ the azimuthal orientation of the  rotation of spiral around axis through Galactic poles \\

$R_\rmn{comp}$ & 7.1 kpc &  & Scale radius of compressed spiral arms.\\

$C_0$ & 2.5 & $c(r)=\left\{ \begin{array}{ll}
  C_0                                 & \mbox{  if $r\le r_\mathrm{cc}$ } \\
  C_0 (r/r_\mathrm{cc})^{-3} & \mbox{  if $r> r_\mathrm{cc}$ } \\ 
\end{array}\right. $ & Arm compression amplitude, tailing off after a region of constant compression.  The contrast, $\epsilon\equiv n^\mathrm{(d)}/n^\mathrm{(u)}$, is the ratio of downstream (post-shock) to upstream (pre-shock) density, in which case $C_0 \approx \epsilon -1$ if we consider arm versus inter-arm instead of down- versus upstream.)  \\

$r_\mathrm{cc}$ & 12 kpc & See $c(r)$. & Region of constant compression.\\

$d_0$ & 0.3 kpc & $d_0(r)=d_0/(c(r)B(r)) $  & Defines the base width
of arm enhancement, which varies with radius.\\

$h_\mathrm{c}$  & 2~kpc  &  $B_\mathrm{comp}(z) = \mathrm{sech}^2(z/h_{\mathrm c})$ & Scale height of the spiral compression. \\

\hline
\multicolumn{4}{c} {\bf Isotropic random and ordered random magnetic fields:}  \\

 \multicolumn{4}{c}{ $\mathbf{B}_\mathrm{iso} + \mathbf{B}_\mathrm{ord} = \left[ \mathbf{B}_\mathrm{GRF} B_\mathrm{GRF}(z) B_\mathrm{GRF}(r) + \sum_i^N \mathbf{B}_{\mathrm{arm},i}^\mathrm{iso} \right]  +  \sum_i^N \mathbf{B}_{\mathrm{arm},i}^\mathrm{ord}$}\\

\hline

$h_{\mathrm rms}$  & 2~kpc  &  $B_\mathrm{GRF}(z) = \mathrm{sech}^2(z/h_{\mathrm rms})$ & Global scale height of the turbulent field field. \\

$R_{\mathrm GRF}$  & 20~kpc  &  $B_\mathrm{GRF}(r) = \exp(-r^2/R^2_{\mathrm GRF})$ & Global scale radius of the turbulent field. \\

$\alpha$ & -0.37 & $P_B(k)\equiv \left< |{\bf
    B}_\rmn{GRF}(k)|^2\right>\propto k^\alpha$ and \hspace*{2.5cm} \linebreak ${\bf
  B}_\rmn{GRF}({\bf x}) = \mathcal{F}^{-1}\left[ {\bf B}_\rmn{GRF}({\bf k}) \right]
  $& Power law spectral
index of initial GRF;  default of -0.37  from \citet{han:2004} in 1D
(compare to 3D Kolmogorov turbulence prediction of $-5/3$). \\

$D_\rmn{co}$ & 0.1 kpc & $B_\rmn{GRF}(k)=0$ for $k<1/D_\rmn{co}$ & Cutoff maximum of GRF fluctuations (minimum determined by resolution) \\

$B_\rmn{rms}$ & 3.5 & $B_\rmn{rms}\equiv \left<B_\rmn{GRF}^2({\bf x})\right>^{1/2}$ & Total RMS amplitude of GRF fluctuations \\

$f_\rmn{ord}$ & 0.15 & ${\bf B}_\mathrm{arm,i}^\mathrm{iso}=\rho_c  {\bf  B}_\rmn{GRF} $ and ${\bf B}_\mathrm{arm,i}^\mathrm{ord}= f_\rmn{ord} \rho_c {\bf B}_\rmn{proj} $  & Relates ordered to isotropic random component. $\bf{B}_\rmn{proj}$ is the component of the GRF parallel to the coherent field direction, i.e. the spiral. \\

\hline
\end{tabular}
\end{minipage}
\end{center}
\caption{Table of magnetic field modelling parameters as described in
  \S~\ref{sec:modeling}.  The total field is the sum of the coherent,
  isotropic random, and ordered random components.  The coherent and
  isotropic random each have a (statistically) axisymmetric and a spiral component,
  while the ordered random only has the spiral component.  Parameter
  values for  the spiral are the same for each component in this work
  with the exception of $C_0$, which in
  Fig.~\ref{fig:profile_simple_comparison} is different for the
  isotropic random component only.}
\label{tab:params_fields}
\end{table*}

\clearpage

\begin{table}
\begin{tabular}{ | l |  l | p{6cm} |}	
\hline
\multicolumn{3}{c}{ {\bf Dust emissivity model parameters:} }\\
\hline
Parameter & Default & Description \\
\hline
$h_\mathrm{g,d}$ & 0.4~kpc  & Scale height of dust uncompressed component. \\
$h_\mathrm{c,d}$ & 0.1~kpc  & Scale height of the dust arms. \\
$R_\mathrm{scale,d}$ & 8~kpc  & Scale radius of the uncompressed component.\\
$R_\mathrm{comp,d}$ & 13~kpc  & Scale radius of the compressed spiral arms.\\
$C_{0,\mathrm{d}}$ & 6  & Compression factor for dust.\\
$d_{0,\mathrm{d}}$ & 0.1~kpc  & Base width of dust arms.  The width
varies with $r$ as described in Table~\ref{tab:params_fields}.\\
\end{tabular}
\caption{Table of parameters for the dust density distribution.  The parametrization is the same as for the magnetic field models defined in Table~\ref{tab:params_fields}, an axisymmetric component plus a spiral compressed component, and any parameters not listed here are the same.}
\label{tab:params_dust}
\end{table}

\label{lastpage}

\end{document}